\documentclass[preprint,aps]{revtex4}
\usepackage{amssymb,epsf}

\begin{document} %updated 13/1/2010%

\title{Thermodynamics of higher dimensional topological\\ charged AdS black branes in dilaton gravity}
\author{S. H. Hendi$^{1,2}$\footnote{email address: hendi@mail.yu.ac.ir},
A. Sheykhi$^{3} $\footnote{email address: sheykhi@mail.uk.ac.ir}
and M. H. Dehghani$^{4} $\footnote{email address:
mhd@shirazu.ac.ir}} \affiliation{$^1$ Physics Department, College
of Sciences, Yasouj University, Yasouj
75914, Iran\\
$^2$ Research Institute for Astronomy and Astrophysics of Maragha (RIAAM),
Maragha, Iran\\
$^3$Department of Physics, Shahid Bahonar University, P.O. Box
76175-132, Kerman, Iran\\
$^4$Physics Department and Biruni Observatory, Shiraz University,
Shiraz 71454, Iran}

\begin{abstract}
In this paper, we study topological AdS black branes of
$(n+1)$-dimensional Einstein-Maxwell-dilaton theory and
investigate their properties. We use the area law, surface gravity
and Gauss law interpretations to find entropy, temperature and
electrical charge, respectively. We also employ the modified Brown
and York subtraction method to calculate the quasilocal mass of
the solutions. We obtain a Smarr-type formula for the mass as a
function of the entropy and the charge, compute the temperature
and the electric potential through the Smarr-type formula and show
that these thermodynamic quantities coincide with their values
which are calculated through using the geometry. Finally, we
perform a stability analysis in the canonical ensemble and
investigate the effects of the dilaton field and the size of black
brane on the thermal stability of the solutions. We find that
large black branes are stable but for small black brane, depending
on the value of dilaton field and type of horizon, we encounter
with some unstable phases.
\end{abstract}

\maketitle

\section{Introduction}

The discovery of a close relationship between the nature of quantum gravity
and the thermodynamics of black holes has been one of the most important
developments in general relativity in the past decades. Strong motivation
for studying thermodynamics of black holes originates from the fact that
they have a very natural thermodynamic description. For example, black holes
have an entropy and temperature related to their horizon area and surface
gravity, respectively, and also one can investigate their thermal stability.
With the appearance of the anti-de Sitter/conformal field theory
correspondence (AdS/CFT) \cite{ADSCFT1}, such black holes in asymptotically
AdS space become even more interesting since one can gain some significant
relations between the thermodynamical properties of the AdS black holes and
the dual conformal field theory \cite{ADSCFT11,ADSCFT2,ADSCFT3}.

On the other hand, it is a general belief that in four dimensions the
topology of the event horizon of an asymptotically flat stationary black
hole is uniquely determined to be the two-sphere $S^{2}$ \cite{Haw1,Haw2}.
Hawking's theorem requires the integrated Ricci scalar curvature with
respect to the induced metric on the event horizon to be positive \cite{Haw1}%
. This condition applied to two-dimensional manifolds determines
uniquely the topology. The ``topological censorship theorem'' of
Friedmann, Schleich and Witt is another indication of the
impossibility of non spherical horizons \cite{FSW1,FSW2}. However,
when the asymptotic flatness of spacetime is violated, there is no
fundamental reason to forbid the existence of static or stationary
black holes with nontrivial topologies. It has been shown that for
asymptotically AdS spacetime, in the four-dimensional
Einstein-Maxwell theory, there exist black hole solutions whose
event horizons may have zero or negative constant curvature and
their topologies are no longer the two-sphere $S^{2}$. The
properties of these black holes are quite different from those of
black holes with usual spherical topology horizon, due to the
different topological structures of the event horizons. Besides,
the black hole thermodynamics is drastically affected by the
topology of the event horizon. It was argued that the Hawking-Page
phase transition \cite{Haw3} for the Schwarzschild-AdS black hole
does not occur for locally AdS black holes whose horizons have
vanishing or negative constant curvature, and they are thermally
stable \cite {Birm}. The studies on the topological black holes
have been carried out extensively in many aspects (see e.g.
\cite{Lemos,Cai2,Bril1,Cai33,Cai44}). In this paper we shall
consider topological black branes in the presence of dilaton and
electromagnetic fields in all higher dimensions. The action of the
$(n+1)$-dimensional $(n\geq 3)$ Einstein-Maxwell-dilaton gravity
can be written as
\begin{equation}
I=-\frac{1}{16\pi }\int d^{n+1}x\sqrt{-g}\left( {R}-K(\Phi )-V(\Phi )+%
\mathcal{L}(\Phi ,F)\right) ,  \label{Action}
\end{equation}
where ${R}$ is the Ricci scalar\textbf{, }$K(\Phi )=4(\nabla \Phi
)^{2}/(n-1) $ is a kinetic term, $V(\Phi )$ is a potential term for the
dilaton field $\Phi $, and $\mathcal{L}(\Phi ,F)=-e^{-4\alpha \Phi
/(n-1)}F_{\mu \nu }F^{\mu \nu }$ is a coupled Lagrangian between scalar
dilaton and electromagnetic fields. In $\mathcal{L}(\Phi ,F)$, $\alpha $ is
an arbitrary constant governing the strength of the coupling between the
dilaton and the Maxwell field, $F_{\mu \nu }=\partial _{\mu }A_{\nu
}-\partial _{\nu }A_{\mu }$ is the electromagnetic field tensor and $A_{\mu }
$ is the electromagnetic potential. While $\alpha =0$ corresponds to the
usual Einstein-Maxwell-scalar theory, $\alpha =1$ indicates the
dilaton-electromagnetic coupling that appears in the low energy string
action in Einstein's frame.

Some attempts have been made to explore various solutions of
Einstein-Maxwell-dilaton gravity. The dilaton field couples in a nontrivial
way to other fields such as gauge fields and results into interesting
solutions for the background spacetime \cite{CDB1,CDB2}. These scalar
coupled black hole solutions \cite{CDB1,CDB2}, however, are all
asymptotically flat. It was argued that with the exception of a pure
cosmological constant, no dilaton-de Sitter or anti-de Sitter black hole
solution exists with the presence of only one Liouville-type dilaton
potential \cite{MW}. In the presence of one or two Liouville-type
potentials, black hole spacetimes which are neither asymptotically flat nor
(A)dS have been explored by many authors (see e.g. \cite
{CHM,Clem,Sheykhi0,Sheykhi1,Hendi2}). Recently, the \textquotedblleft
cosmological constant term\textquotedblright\ in the dilaton gravity has
been found by Gao and Zhang \cite{Gao}. With an appropriate combination of
three Liouville-type dilaton potentials, they obtained the static dilaton
black hole solutions which are asymptotically (A)dS in four and higher
dimensions. In such a scenario AdS spacetime constitutes the vacuum state
and the black hole solution in such a spacetime becomes an important area to
study \cite{ADSCFT1}. For an arbitrary value of $\alpha $ in AdS spaces the
form of the dilaton potential ${V}({\Phi })$ in $n+1$ dimensions is chosen
as \cite{Gao}
\begin{eqnarray}
{V}({\Phi }) &=&\frac{2\Lambda }{n(n-2+\alpha ^{2})^{2}}\Bigg\{-\alpha ^{2}%
\left[ (n+1)^{2}-(n+1)\alpha ^{2}-6(n+1)+\alpha ^{2}+9\right] e^{-4(n-2){%
\Phi /}[(n-1)\alpha ]}  \nonumber \\
&&+(n-2)^{2}(n-\alpha ^{2})e^{4\alpha {\Phi /}(n-1)}+4\alpha
^{2}(n-1)(n-2)e^{-2{\Phi }(n-2-\alpha ^{2})/[(n-1)\alpha ]}\Bigg \},
\label{V(Phi)}
\end{eqnarray}
where $\Lambda $ is the cosmological constant. The motivations for
studying such dilaton black holes with nonvanishing cosmological
constant originate from supergravity theory. Gauged supergravity
theories in various dimensions are obtained with negative
cosmological constant in a supersymmetric theory. In addition, it
has been shown that one may consider a Big Bang model of the
Universe in the presence of dilaton field and presented
Liouville-type potential which can mimic the matter (including
dark matter) and dark energy. This model predict age of the
Universe, transition redshift, Big Bang nucleosynthesis and
evolution of dark energy agree with current observations
\cite{GZH}. Also, this type of potential can be obtained when a
higher dimensional theory is compactified to four dimensions,
including various super gravity models \cite{RaduTch} (see also
\cite{Giddings} for a recent discussion of these aspects). In
particular, for special values of coupling constant, $\alpha$,
this potential reduce to the supersymmetry potential of Gates and
Zwiebach in string theory \cite{RaduTch}.

For later convenience we redefine $\Lambda =-n(n-1)/2l^{2}$, where $l$ is
the AdS radius of spacetime. It is clear the cosmological constant is
coupled to the dilaton in a very nontrivial way. In the absence of the
dilaton field action (\ref{Action}) reduces to the action of
Einstein-Maxwell gravity with cosmological constant. Considering this type
of dilaton potential, one can extract successfully the AdS solutions of
Einstein--Maxwell-dilaton gravity \cite{Sheykhi2,SheDehHen2010}.

The rest of this paper is outlined as follows. In the next section, we
consider the field equations of Einstein-Maxwell-dilaton gravity and present
the $(n+1)$-dimensional topological AdS black brane solutions and investigate
their properties. In section \ref{Therm}, we obtain the conserved and
thermodynamic quantities of the solutions and verify the validity of the
first law of black brane thermodynamics. We perform a stability analysis in
the canonical ensemble and disclose the effect of the dilaton field on the
thermal stability of the solutions in section \ref{Stab}. Conclusions are
drawn in the last section.

\section{Field equations and solutions\label{Field}}

Varying action (\ref{Action}) with respect to the metric tensor $g_{\mu \nu
} $, the dilaton field $\Phi $ and the electromagnetic potential $A_{\mu }$,
the equations of motion are obtained as
\begin{equation}
{R}_{\mu \nu }=\frac{4}{n-1}\left( \partial _{\mu }\Phi \partial _{\nu }\Phi
+\frac{1}{4}g_{\mu \nu }V(\Phi )\right) +2e^{-4\alpha \Phi /(n-1)}F_{\mu
\eta }F_{\nu }^{\text{ }\eta }-\frac{g_{\mu \nu }}{(n-1)}\mathcal{L}(\Phi
,F),  \label{FE1}
\end{equation}
\begin{equation}
\nabla ^{2}\Phi =\frac{n-1}{8}\frac{\partial V}{\partial \Phi }-\frac{\alpha
}{2}\mathcal{L}(\Phi ,F),  \label{FE2}
\end{equation}
\begin{equation}
\nabla _{\mu }\left( e^{-{4\alpha \Phi }/({n-1})}F^{\mu \nu }\right) =0.
\label{FE3}
\end{equation}
Here we want to obtain the $(n+1)$-dimensional static solutions of Eqs. (\ref
{FE1}), (\ref{FE2}) and (\ref{FE3}). We assume that the metric has the
following form
\begin{equation}
d{s}^{2}=-N^{2}(\rho )f^{2}(\rho )dt^{2}+\frac{d\rho ^{2}}{f^{2}(\rho )}%
+\rho ^{2}{R^{2}(\rho )}d\Omega _{n-1}^{2},  \label{metric}
\end{equation}
where
\[
d\Omega _{n-1}^{2}=\left\{
\begin{array}{cc}
d\theta _{1}^{2}+\sinh ^{2}\theta _{1}d\theta _{2}^{2}+\sinh ^{2}\theta
_{1}\sum\limits_{i=3}^{n-1}\prod\limits_{j=2}^{i-1}\sin ^{2}\theta
_{j}d\theta _{i}^{2} & k=-1 \\
\sum\limits_{i=1}^{n-1}d\phi _{i}^{2} & k=0
\end{array}
\right.
\]
represents the line element of an $(n-1)$-dimensional hypersurface with
constant curvature $(n-1)(n-2)k$. It is notable
that positive curvature horizon ($k=1$) has been investigated in \cite
{SheDehHen2010}. Here $N(\rho )$, $f(\rho )$ and $R(\rho )$ are functions of
$\rho $ which should be determined. First of all, the Maxwell equations (\ref
{FE3}) can be integrated immediately, where all the components of ${F}_{\mu
\nu }$ are zero except ${F}_{t\rho }$%
\begin{equation}
{F}_{t\rho }=N(\rho )\frac{qe^{4\alpha {\Phi /(n-1)}}}{\left( \rho R\right)
^{n-1}},  \label{Ftr}
\end{equation}
where $q$, an integration constant, is the charge parameter of the black
brane. Our aim here is to construct exact, $(n+1)$-dimensional topological
AdS black brane solutions of Eqs. (\ref{FE1})-(\ref{FE3}), with the dilaton
potential (\ref{V(Phi)}) for an arbitrary dilaton coupling parameter $\alpha
$ and investigate their properties. Using metric (\ref{metric}) and the
Maxwell field (\ref{Ftr}), one can show that the system of equations (\ref
{FE1})-(\ref{FE2}) have solutions of the form
\begin{eqnarray}
N^{2}(\rho ) &=&\Upsilon ^{-\gamma (n-3)},  \label{Nrho} \\
f^{2}(\rho ) &=&\frac{\rho ^{2}}{l^{2}}\Upsilon ^{(n-2)\gamma }+\left[
k-\left( \frac{c}{\rho }\right) ^{n-2}\right] \Upsilon ^{1-\gamma },
\label{frho} \\
{\Phi }(\rho ) &=&\frac{n-1}{4}\sqrt{\gamma (2+2\gamma -n\gamma )}\ln
\Upsilon ,  \label{Phirho} \\
R^{2}(\rho ) &=&\Upsilon ^{\gamma },  \label{Rrho} \\
\Upsilon &=&1-\left( \frac{b}{\rho }\right) ^{n-2}.  \nonumber
\end{eqnarray}
Here $c$ and $b$ are integration constants and the constant $\gamma $ is
\begin{equation}
\gamma =\frac{2\alpha ^{2}}{(n-2)(n-2+\alpha ^{2})}.  \label{gamma}
\end{equation}
The charge parameter $q$ is related to $b$ and $c$ by
\begin{equation}
q^{2}=\frac{(n-1)(n-2)^{2}}{2(n-2+\alpha ^{2})}c^{n-2}b^{n-2}.  \label{q}
\end{equation}
For $\alpha \neq 0$ the solutions are not real for $0<\rho <b$ and therefore
we should exclude this region from the spacetime. For this purpose we
introduce the new radial coordinate $r$ as
\begin{equation}
r^{2}=\rho ^{2}-b^{2}\Rightarrow d\rho ^{2}=\frac{r^{2}}{r^{2}+b^{2}}dr^{2}.
\label{Transformation}
\end{equation}
With this new coordinate, the above metric becomes
\begin{equation}
d{s}^{2}=-N^{2}(r)f^{2}(r)dt^{2}+\frac{r^{2}dr^{2}}{(r^{2}+b^{2})f^{2}(r)}%
+(r^{2}+b^{2}){R^{2}(r)}d\Omega _{n-1}^{2},  \label{Metric2}
\end{equation}
where the coordinates $r$ assumes the values $0\leq r<\infty $, and $%
N^{2}(r) $, $f^{2}(r)$, $\Phi (r)$ and $R^{2}(r)$ are now given as
\begin{eqnarray}  \label{gr}
N^{2}(r) &=&\Gamma ^{-\gamma (n-3)},  \label{Nr} \\
f^{2}(r) &=&\frac{r^{2}+b^{2}}{l^{2}}\Gamma ^{(n-2)\gamma }+\left[ k-\left(
\frac{c}{\sqrt{r^{2}+b^{2}}}\right) ^{n-2}\right] \Gamma ^{1-\gamma },
\label{fr} \\
{\Phi }(r) &=&\frac{n-1}{4}\sqrt{\gamma (2+2\gamma -n\gamma )}\ln \Gamma ,
\label{Phir} \\
R^{2}(r) &=&\Gamma ^{\gamma },  \label{Rr} \\
\Gamma &=&1-\left( \frac{b}{\sqrt{r^{2}+b^{2}}}\right) ^{n-2}.  \nonumber
\end{eqnarray}

\subsection{Properties of the solutions:}

It is notable to mention that these solutions are valid for all values of $%
\alpha $. When ($\alpha =0=\gamma $), these solutions describe the $(n+1)$%
-dimensional asymptotically AdS Reissner-Nordstrom black branes. For $b=0$ $%
(\Gamma =1)$, the charge parameter $q$ and the scalar field ${\Phi }(r)$
vanish and our solutions reduce to the solutions of Einstein gravity in the
presence of cosmological constant.

\begin{figure}[tbp]
\epsfxsize=8cm \centerline{\epsffile{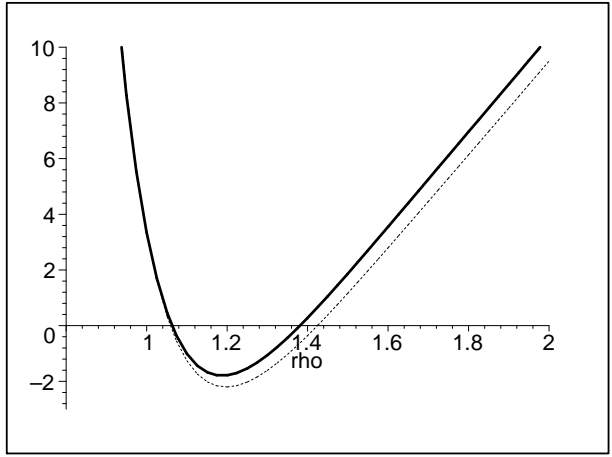}}
\caption{$f^{2}(\protect\rho )$ versus $\protect\rho $ for $b=1$, $c=1.5$, $%
l=0.5$, $n=5$ and $\protect\alpha =0$ (timelike singularity). $k=0$ (bold
line), and $k=-1$ (dashed line).}
\label{Figf21}
\end{figure}

\begin{figure}[tbp]
$
\begin{array}{cc}
\epsfxsize=8cm \epsffile{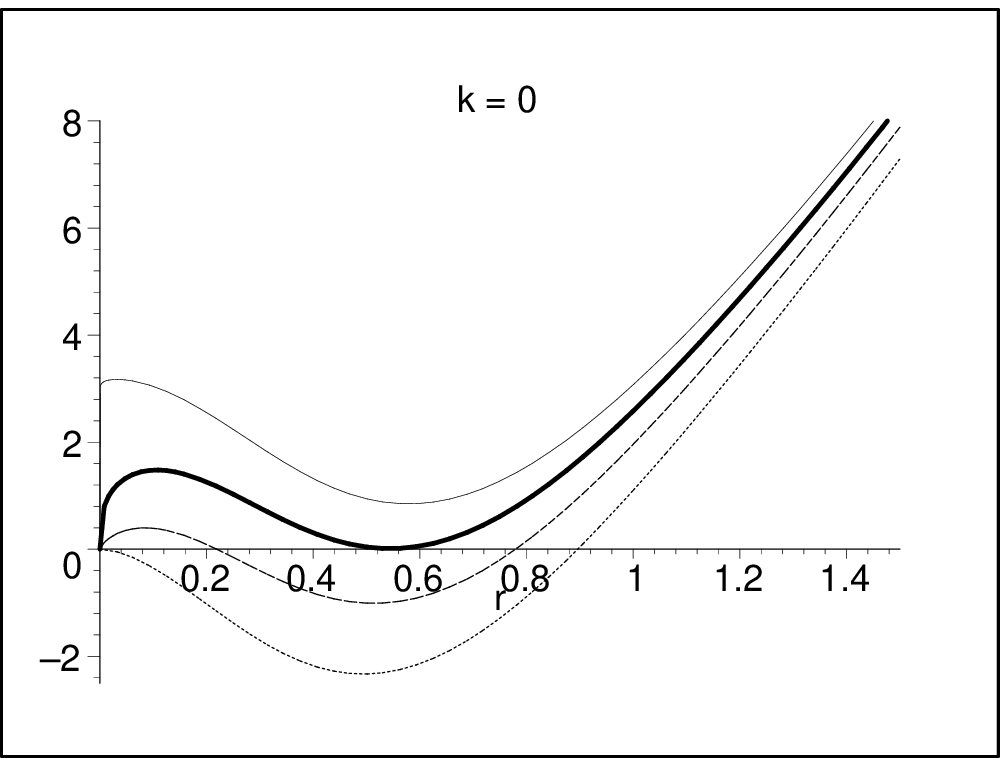} & \epsfxsize=8cm \epsffile{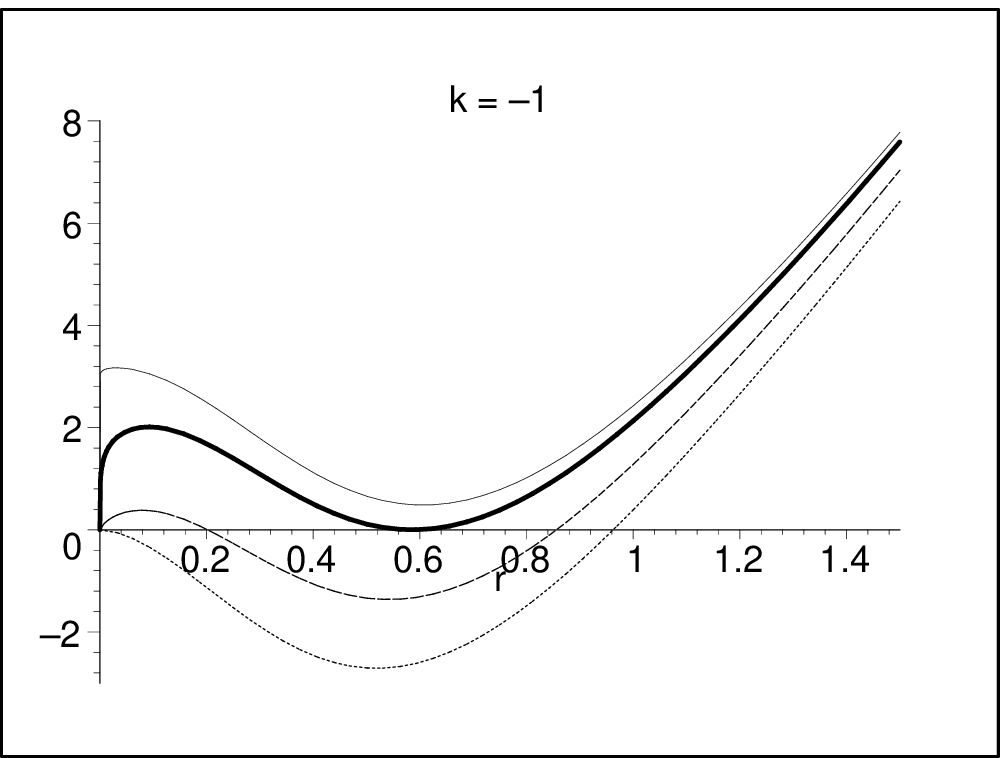}
\end{array}
$%
\caption{$f^{2}(r)$ versus $r$ for $b=1$, $c=2.5$, $l=0.5$ and $n=5$. $%
\protect\alpha =0.1$ (solid line: naked singularity), $\protect\alpha =%
\protect\alpha _{ext}$ (bold line: extreme black brane), $\protect\alpha =0.8$
(dashed line: null singularity with two horizons), and $\protect\alpha =1.2$
(dotted line: null singularity with one horizon), where $\protect\alpha
_{ext}=0.59,0.67$ for $k=-1,0$, respectively.}
\label{Figf221}
\end{figure}

Here we should discuss singularity(ies). After some algebraic
manipulation, one can show that the Kretschmann and the Ricci scalars in $%
(n+1)$-dimensions are finite for $r\neq 0$, but in the vicinity of $r=0$, we
have
\begin{eqnarray*}
R_{\mu \nu \lambda \kappa }R^{\mu \nu \lambda \kappa } &\propto &r^{-4(n-2)
\left[ 1+\left( \alpha ^{2}+n-2\right) ^{-1}\right] }, \\
R &\propto &r^{-2(n-2)\left[ 1+\left( \alpha ^{2}+n-2\right) ^{-1}\right] },
\end{eqnarray*}
and thus they diverge at $r=0$ ($\rho =b$). Thus, $r=0$ is a curvature
singularity. It is worthwhile to note that the scalar field $\Phi (r)$ and
the electromagnetic field $F_{t\rho }$ become zero as $r\rightarrow \infty
(\rho \rightarrow \infty )$. One should note that for nonzero $\alpha$ the
singularity may be null, while it is timelike for $\alpha =0$ (see Figs. \ref
{Figf21} and \ref{Figf221}). Also, figure \ref{Figf221} shows that depending
on the metric parameters, these real solutions may be interpreted as black
brane solutions with inner and outer horizons, an extreme black brane or naked
singularity.
\vspace{0.2cm}

\section{Thermodynamics of AdS dilaton black brane\label{Therm}}

In this section we intend to study thermodynamics of topological dilaton
black branes in the background of AdS spaces. First of all we focus on
entropy. The entropy of the dilaton black hole typically satisfies the
so-called area law of the entropy which states that the entropy of the black
hole is one-quarter of the event horizon area \cite{Beck}. This near
universal law applies to almost all kinds of black objects, including
dilaton black holes, in Einstein gravity \cite{hunt}. It is a matter of
calculation to show that the entropy of the dilaton black brane per unit volume is
\begin{equation}
{S}=\frac{b^{n-1}\Gamma _{+}^{\gamma (n-1)/2}}{4\left( 1-\Gamma
_{+}\right) ^{(n-1)/(n-2)}},  \label{entropy}
\end{equation}
where $\Gamma _{+}=\Gamma (r=r_{+})$. It is notable to mention that in
contrast with the higher derivative gravities that may lead to negative
entropy \cite{negativeS}, the presented entropy is positive definite ($%
0\leq \Gamma _{+}\leq 1$). The Hawking temperature of the dilaton black brane
on the outer horizon $r_{+}$, may be obtained through the use of the
definition of surface gravity,
\begin{equation}
T_{+}=\frac{1}{2\pi }\sqrt{-\frac{1}{2}\left( \nabla _{\mu }\chi _{\nu
}\right) \left( \nabla ^{\mu }\chi ^{\nu }\right) }=\sqrt{r_{+}^{2}+b^{2}}%
\left. \frac{\left( N^{2}f^{2}\right) ^{^{\prime }}}{4\pi Nr}\right|
_{r=r_{+}},  \label{T}
\end{equation}
where $\chi =\partial _{t}$ is the killing vector and a prime stands for the
derivative with respect to $r$. Finding the radius of outer horizon in terms
of the parameters of the metric function is not possible analytically, and
therefore we obtain the constant $c$ in terms of $b$, $\alpha $ and $r_{h}$
by solving $f(r_{h})=0$, where $r_{h}$ is the radius of inner or outer
horizon of the black brane. Substituting $c$ into Eq. (\ref{T}), one obtains
\begin{equation}
{T}_{h}=\frac{b(n-2)\Gamma _{h}^{1-\gamma (n-1)/2}}{4\pi \left( 1-\Gamma
_{h}\right) ^{1/(n-2)}}\Bigg\{\frac{\left[ \gamma (n-1)-1\right] }{%
l^{2}\Gamma _{h}^{2-\gamma (n-1)}}-\frac{(n-1)\left[ \gamma (n-2)-2\right] }{%
(n-2)l^{2}\Gamma _{h}^{1-\gamma (n-1)}}+\frac{k(1-\Gamma _{h})^{2/(n-2)}}{%
b^{2}}\Bigg\},  \label{Tem}
\end{equation}
where $\Gamma _{h}=\Gamma (r=r_{h})$.
\begin{figure}[tbp]
$
\begin{array}{cc}
\epsfxsize=8cm \epsffile{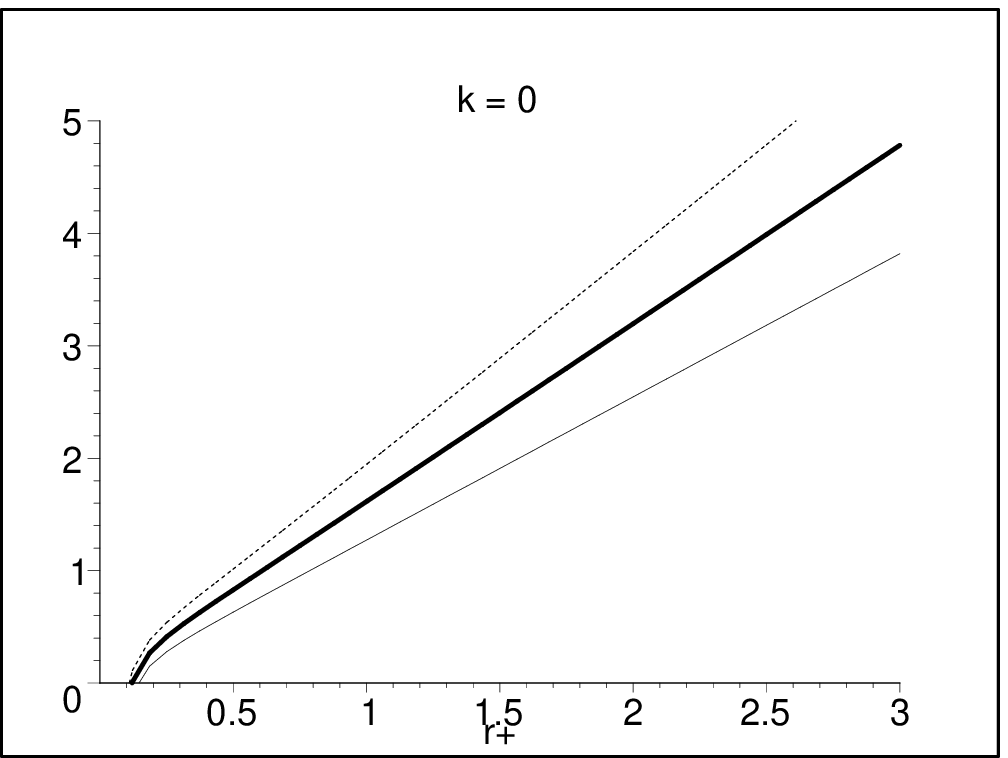} & \epsfxsize=8cm %
\epsffile{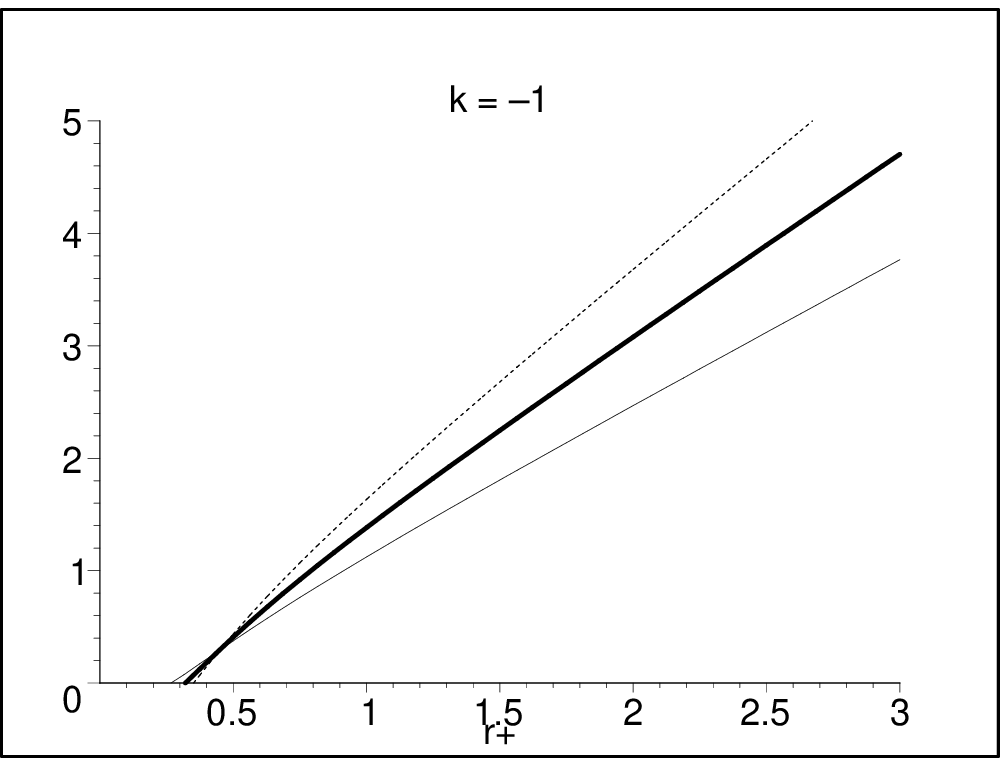}
\end{array}
$%
\caption{T versus $r_{+}$ for $b=0.2$, $l=0.5$ and $\protect\alpha =0.1,$ $%
n=4$ (solid line), $n=5$ (bold line), and $n=6$ (dashed line).}
\label{Fig1}
\end{figure}
\vspace{0.2cm}
\begin{figure}[tbp]
$
\begin{array}{cc}
\epsfxsize=8cm \epsffile{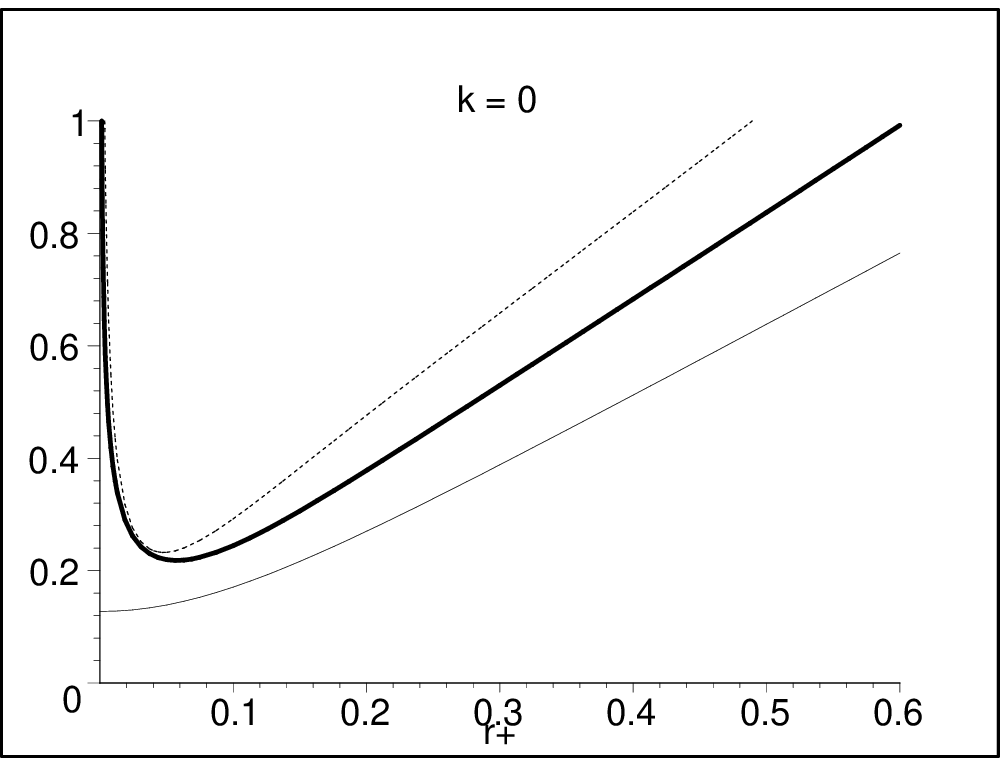} & \epsfxsize=8cm %
\epsffile{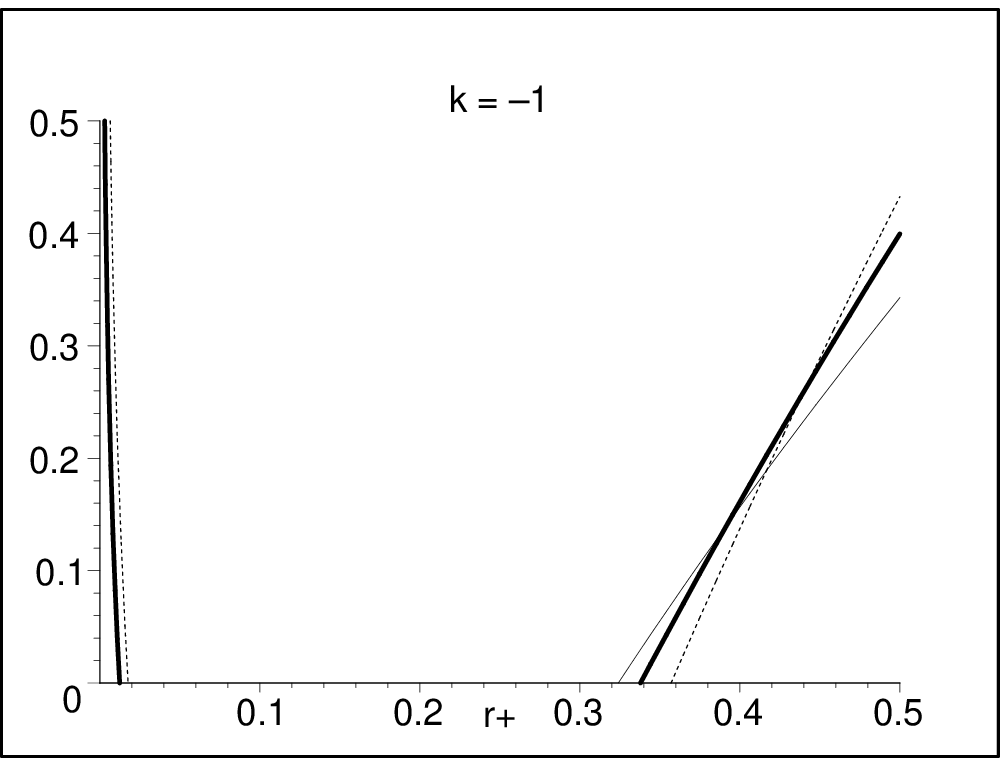}
\end{array}
$%
\caption{T versus $r_{+}$ for $b=0.2$, $l=0.5$ and $\protect\alpha =2,$ $n=4$
(solid line), $n=5$ (bold line), and $n=6$ (dashed line).}
\label{Fig11}
\end{figure}
\vspace{0.2cm}
\begin{figure}[tbp]
$
\begin{array}{cc}
\epsfxsize=8cm \epsffile{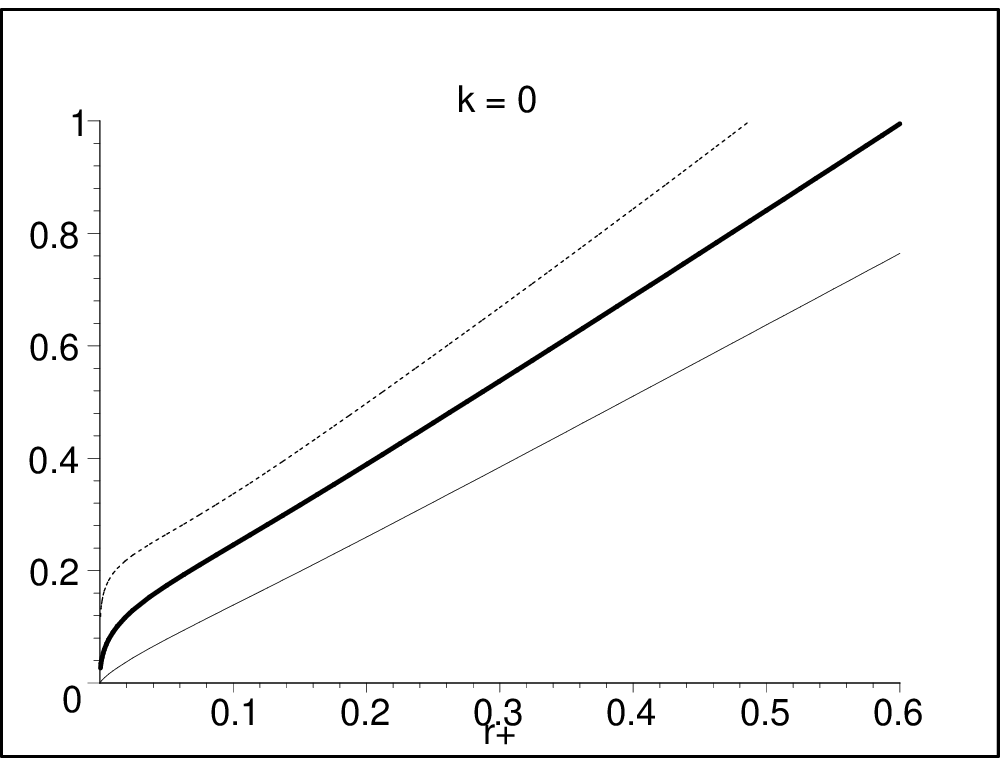} & \epsfxsize=8cm %
\epsffile{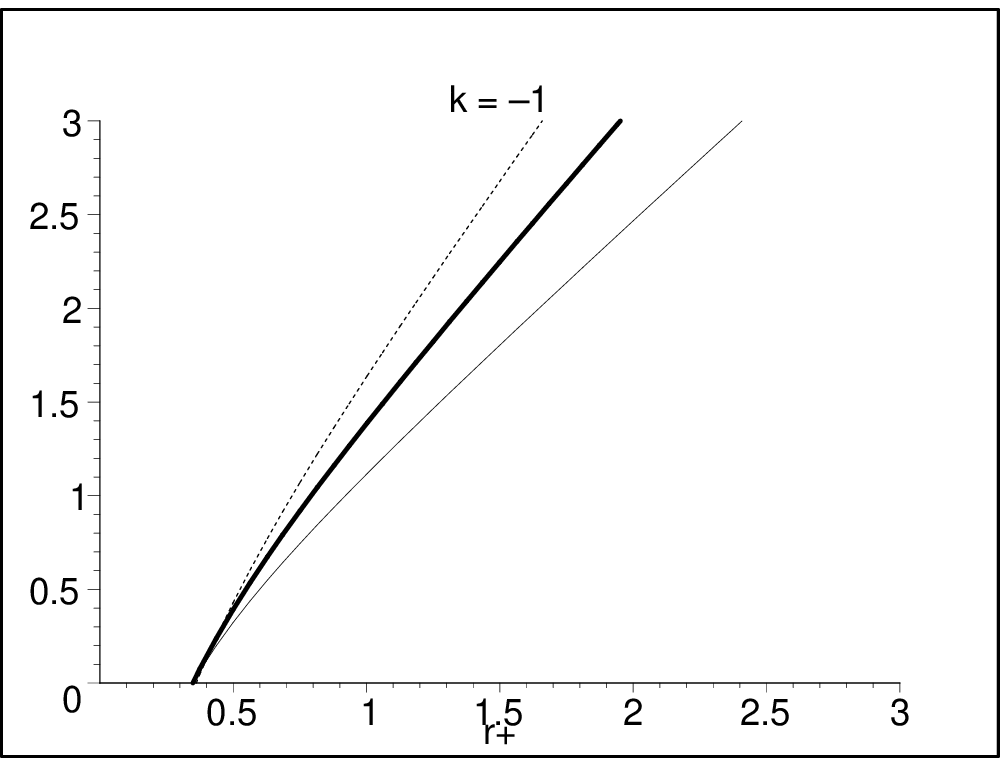}
\end{array}
$%
\caption{T versus $r_{+}$ for $b=0.2$, $l=0.5$ and $\protect\alpha =5,$ $n=4$
(solid line), $n=5$ (bold line), and $n=6$ (dashed line).}
\label{Fig111}
\end{figure}
\vspace{0.2cm}
\begin{figure}[tbp]
$
\begin{array}{cc}
\epsfxsize=8cm \epsffile{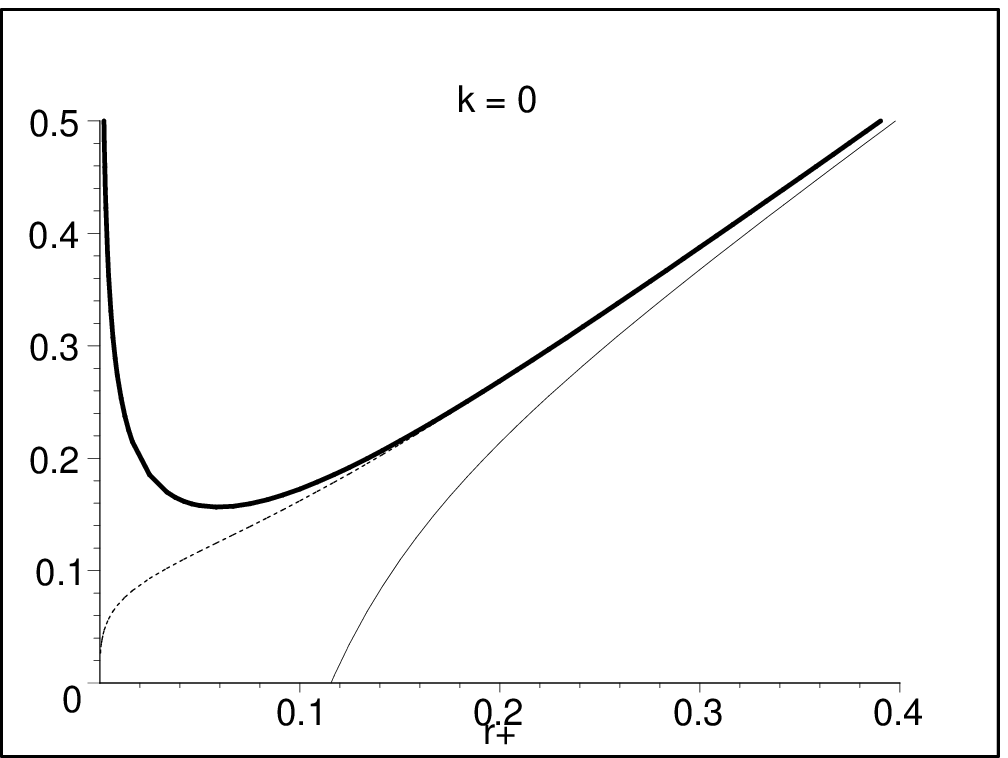} & \epsfxsize=8cm %
\epsffile{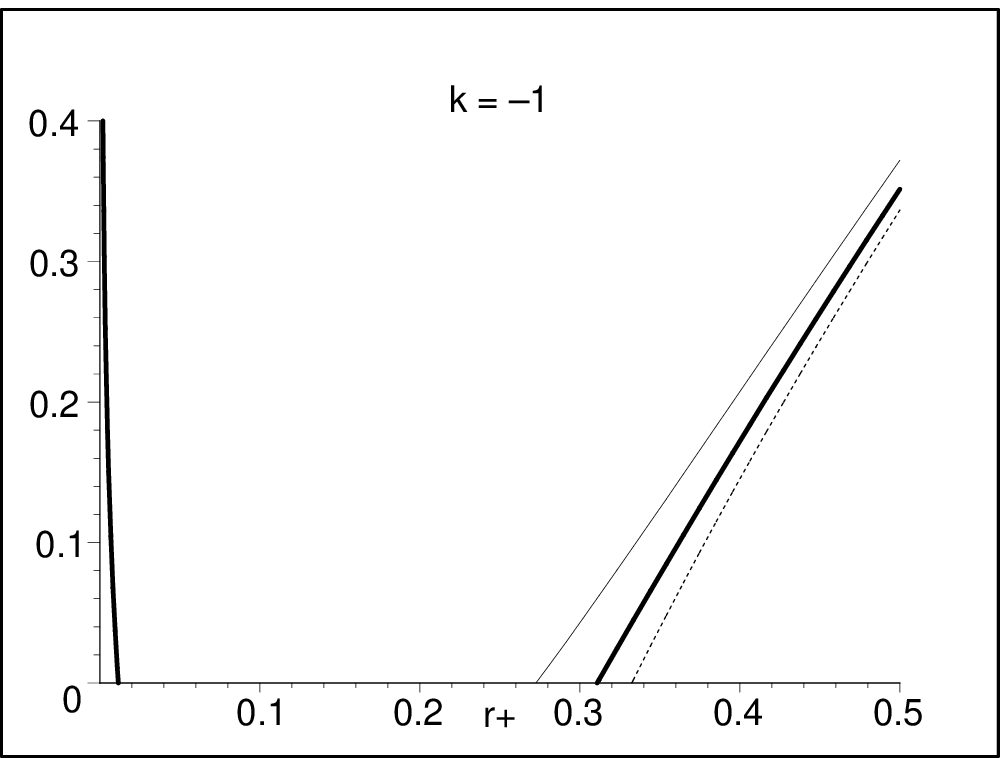}
\end{array}
$%
\caption{T versus $r_{+}$ for $b=0.2$, $l=0.5$ and $n=4$, $\protect\alpha
=0.5$ (solid line), $\protect\alpha =1.5$ (bold line), and $\protect\alpha
=2.5$ (dashed line).}
\label{Fig1111}
\end{figure}
The equation $T_{h}=0$ has
one real root for $k=0$:
\begin{equation}
r_{\mathrm{ext}}=b\left\{ \left( \frac{(n-1)\left[ 2-\gamma (n-2)\right] }{n}%
\right) ^{2/(n-2)}-1\right\} ^{1/2},  \label{rext}
\end{equation}
while it may have two real roots ($r_{\mathrm{1ext}}$\ and $r_{\mathrm{2ext}}
$) for $k=-1$. These roots are the radius of the extreme black branes. We are
interested in the thermodynamics of event horizon, $r_{+}$, of the black
branes and therefore we consider $T_{+}\geq 0$. The negative values of $T_{h}$%
\ associated to the temperature of inner horizon. The radius of event
horizon $r_{+}\geq r_{\mathrm{ext}}$\ for $k=0$, and $r_{+}\leq r_{\mathrm{%
1ext}}$\ or $r_{+}\geq r_{2\mathrm{ext}}$\ for $k=-1$. These facts can be
seen in Figs. \ref{Fig1} - \ref{Fig1111}, which show the temperature versus $%
r_{+}$\ in various dimensions. In order to be more clear, we discuss these
figures for $k=0$\ and $k=-1$, separately.

\textbf{$k=0$}:

As one can see in Figs. \ref{Fig1} - \ref{Fig1111}, there exist extreme
black brane with radius $r_{\mathrm{ext}}$\ provided $\alpha <\alpha _{%
\mathrm{ext}}$, where
\begin{equation}
\alpha _{\mathrm{ext}}^{2}=\frac{n-2}{n}\left( n-2+2(n-1)\left[ \left( 1+%
\frac{r_{\mathrm{ext}}^{2}}{b^{2}}\right) ^{1-n/2}-1\right] \right) .
\label{alphaCrit}
\end{equation}
That is the dilaton field removes the existence of extreme black branes.

\textbf{$k=-1$}:

For black branes with hyperbolic horizon and medium values of
$\alpha $, numerical analysis shows that the equation $T_{h}=0$ \
has two real roots and therefore one can have both small
(r$_{+}\leq r_{\mathrm{1ext}}$) and large (r$_{+}\geq
r_{2\mathrm{ext}}$) black branes. This can be seen on Figs.
\ref{Fig1} - \ref{Fig1111}. Concerning the metric function $f(r)$
and temperature $T$, Figs. \ref{Figf221} - \ref{Fig1111}, one can
find that for small values of $\alpha$ they behave like
charged-AdS black branes while for large values of coupling
constant $\alpha$, they are approximately Schwarzschild-AdS black
branes.

Inserting solutions (\ref{frho})-(\ref{Rrho}) in Eq. (\ref{Ftr}), with
considering the new coordinate (\ref{Transformation}), the electromagnetic
field can be simplified as
\begin{equation}
F_{tr}=\frac{q}{\left( r^{2}+b^{2}\right) ^{\left( n-1\right) /2}}.
\label{Ftrho2}
\end{equation}
As one can see from Eq. (\ref{Ftrho2}), in the background of AdS universe,
the dilaton field does not exert any direct influence on the matter field $%
F_{tr}$, however, the dilaton field modifies the geometry of the spacetime
as it participates in the field equations. This is in contrast to the
solutions presented in \cite{CHM,Clem,Sheykhi0,Sheykhi1}. The solutions of
Ref. \cite{CHM,Clem,Sheykhi0,Sheykhi1} are neither asymptotically flat nor
(A)dS and the gauge field crucially depends on the scalar dilaton field.

The electric charge of the black brane per unit volume, $Q$, can be found by calculating the
flux of the electromagnetic field at infinity (Gauss theorem), obtaining
\begin{equation}
Q=\frac{1}{4\pi }\int_{\rho \rightarrow \infty }d^{n-1}x\sqrt{-{g}}{F}%
_{t\rho }=\frac{\Omega _{n-1}}{4\pi }q.  \label{Q}
\end{equation}
Let us return to Eq. (\ref{Ftrho2}). The gauge potential $A_{t}$
corresponding to the electromagnetic field (\ref{Ftrho2}) can be easily
obtained as
\[
A_{t}=-\frac{q}{(n-2)\left( r^{2}+b^{2}\right) ^{\left( n-2\right) /2}}.
\]
The electric potential $U$, measured at infinity with respect to the
horizon, is defined by \cite{Cal}
\begin{equation}
U=A_{\mu }\chi ^{\mu }\left| _{r\rightarrow \infty }-A_{\mu }\chi ^{\mu
}\right| _{r=r_{+}},  \label{Pot}
\end{equation}
where $\chi =\partial _{t}$ is the null generator of the horizon. Therefore,
the electric potential may be obtained as
\begin{equation}
U=\frac{q}{(n-2)\left( r_{+}^{2}+b^{2}\right) ^{\left( n-2\right) /2}}.
\label{pot}
\end{equation}
The quasilocal mass of the dilaton AdS black hole can be calculated through
the use of the subtraction method of modified Brown and York (BY) \cite{BY}. Such a
procedure causes the resulting physical quantities to depend on the choice
of reference background. In order to use the BY method the metric should
have the form
\begin{equation}
ds^{2}=-W({\mathcal{R}})dt^{2}+\frac{d\mathcal{R}^{2}}{V(\mathcal{R})}+%
\mathcal{R}^{2}d\Omega ^{2}.  \label{Mets}
\end{equation}
Thus, we should write the metric (\ref{metric}) in the above form. To do
this, we perform the following transformation \cite{DB}:
\[
\mathcal{R}=\rho \Upsilon ^{\gamma /2}.
\]
It is a matter of calculations to show that the metric (\ref{metric}) may be
written as (\ref{Mets}) with the following $W$ and $V$:
\begin{eqnarray*}
W(\mathcal{R}) &=&N^{2}(\rho (\mathcal{R}))f^{2}(\rho (\mathcal{R})), \\
V(\mathcal{R)} &=&f^{2}(\rho (\mathcal{R}))\left( \frac{d\mathcal{R}}{d\rho }%
\right) ^{2}=\Upsilon ^{(\gamma -2)}\left[ 1+\frac{1}{2}\left( \gamma
(n-2)-2\right) \left( 1-\Upsilon \right) \right] ^{2}f^{2}(\rho (\mathcal{R}%
)).
\end{eqnarray*}
The background metric is chosen to be the metric (\ref{Mets}) with
\begin{equation}
W_{0}(\mathcal{R})=V_{0}(\mathcal{R})=f_{0}^{2}(\rho (\mathcal{R}))=k+\frac{%
\rho ^{2}}{l^{2}}+\left\{
\begin{array}{cc}
-\frac{2b\rho \alpha ^{2}}{(1+\alpha ^{2})l^{2}}+\frac{b^{2}\alpha ^{4}}{%
(1+\alpha ^{2})^{2}l^{2}} & \text{for \ }n=3 \\
-\frac{b^{2}\alpha ^{2}}{(2+\alpha ^{2})l^{2}} & \text{for \ }n=4 \\
0 & \text{for \ }n\geqslant 5
\end{array}
\right.  \label{W0}
\end{equation}
As one can see from the above equation, the solutions for $n=3$
and $n=4$ have not ''exact'' asymptotic AdS behavior. Because of
this point, we cannot use the AdS/CFT correspondence to compute
the mass. Indeed, for $n\geqslant 5 $ the metric is exactly
asymptotically AdS, while for $n=3,4$ it is \textit{approximately}
asymptotically AdS. This is due to the fact that if one computes
the Ricci scalar then it is not equal to $-n(n+1)/l^{2}$. It is
well-known that the Ricci scalar for AdS spacetime should have
this value (see e.g. \cite{Weinberg}). Also, the metrics with
$f_{0}^{2}(\rho )$ given by Eq. (\ref{W0}) for $n=3$ and $n=4$ do
not satisfy the Einstein equation with the cosmological constant,
while an AdS spacetime should satisfy the
Einstein equation with cosmological constant. On the other side, at large $%
\rho $, the metric behaves as $\rho ^{2}$ in all dimensions and therefore we
used the word \textit{''approximately''} asymptotically AdS.

To compute the conserved mass of the spacetime, we choose a timelike Killing
vector field $\xi $ on the boundary surface $\mathcal{B}$ of the spacetime (%
\ref{Mets}). Then the quasilocal conserved mass can be written as
\begin{equation}
\mathcal{M}=\frac{1}{8\pi }\int_{\mathcal{B}}d^{2}\varphi \sqrt{\sigma }%
\left\{ \left( K_{ab}-Kh_{ab}\right) -\left(
K_{ab}^{0}-K^{0}h_{ab}^{0}\right) \right\} n^{a}\xi ^{b},
\end{equation}
where $\sigma $ is the determinant of the metric of the boundary $\mathcal{B}
$, $K_{ab}^{0}$ is the extrinsic curvature of the background metric and $%
n^{a}$ is the timelike unit normal vector to the boundary $\mathcal{B}$. In
the context of counterterm method, the limit in which the boundary $\mathcal{%
B}$ becomes infinite ($\mathcal{B}_{\infty }$) is taken, and the counterterm
prescription ensures that the action and conserved charges are finite.
Although the explicit function $f(\rho (\mathcal{R}))$ cannot be obtained,
but at large $\mathcal{R}$ this can be done. Thus, one can calculate the
mass per unit volume through the use of the above modified Brown and York formalism as
\begin{equation}
{M}=\frac{n-1}{16\pi }\left[ c^{n-2}+k\left( \frac{n-2-\alpha
^{2}}{n-2+\alpha ^{2}}\right) b^{n-2}\right] .  \label{Mass}
\end{equation}
In the absence of a non-trivial dilaton field ($\alpha =0$), this expression
for the mass reduces to the mass of the $(n+1)$-dimensional asymptotically
AdS black brane.

Finally, we check the first law of thermodynamics for the black brane. In
order to do this, we obtain the mass $M$ as a function of extensive
quantities $S$ and $Q$. Using the expression for the charge, the mass and
the entropy given in Eqs. (\ref{Q}), (\ref{Mass}) and (\ref{entropy}), we
can obtain a Smarr-type formula per unit volume as
\begin{equation}
M(S,Q)=\frac{(n-1)}{16\pi }\left[ \frac{32\pi ^{2}(n-2+\alpha
^{2})Q^{2}b^{2-n}}{(n-1)(n-2)^{2}}+k\left( \frac{n-2-\alpha ^{2}}{n-2+\alpha
^{2}}\right) b^{n-2}\right] ,  \label{Msmarr}
\end{equation}
where $b=b(Q,S)$. One may then regard the parameters $S$ and $Q$ as a
complete set of extensive parameters for the mass $M(S,Q)$ and define the
intensive parameters conjugate to $S$ and $Q$. These quantities are the
temperature and the electric potential
\begin{eqnarray}
T &=&\left( \frac{\partial {M}}{\partial {S}}\right) _{Q}=\frac{\left( \frac{%
\partial {M}}{\partial {b}}\right) _{Q}\left( \frac{\partial {b}}{\partial {r%
}_{+}}\right) _{Q}}{\left( \frac{\partial {S}}{\partial {r}_{+}}\right)
_{Q}+\left( \frac{\partial {S}}{\partial {b}}\right) _{Q}\left( \frac{%
\partial {b}}{\partial {r}_{+}}\right) _{Q}},  \label{inte1} \\
U &=&\left( \frac{\partial {M}}{\partial {Q}}\right) _{S}=\left( \frac{%
\partial {M}}{\partial {Q}}\right) _{S}+\left( \frac{\partial {M}}{\partial {%
b}}\right) _{S}\left( \frac{\partial {Q}}{\partial {b}}\right) _{S},
\label{inte2}
\end{eqnarray}
where
\begin{eqnarray}
&&\left( \frac{\partial {b}}{\partial {r}_{+}}\right) _{Q}=-\frac{\left(
\frac{\partial {Z}}{\partial {r}_{+}}\right) _{Q}}{\left( \frac{\partial {Z}%
}{\partial {b}}\right) _{Q}}, \\
&&Z=\left[ k-\frac{32\pi ^{2}(n-2+\alpha ^{2})Q^{2}}{(n-1)(n-2)^{2}b^{2n-4}}%
\left( \frac{b}{r_{+}}\right) ^{n-2}\right] \left[ 1-\left( \frac{b}{r_{+}}%
\right) ^{n-2}\right] ^{1-\gamma \left( n-2\right) }  \nonumber \\
&&-\frac{r_{+}^{2}}{l^2}\left[ 1-\left( \frac{b}{r_{+}}\right) ^{n-2}\right]
^{\gamma }.
\end{eqnarray}
Straightforward calculations show that the intensive quantities calculated
by Eqs. (\ref{inte1}) and (\ref{inte2}) coincide with Eqs. (\ref{Tem}) and (%
\ref{pot}). Thus, these thermodynamics quantities satisfy the first law of
black brane thermodynamics,
\begin{equation}
dM=TdS+Ud{Q}.
\end{equation}

\vspace{0.2cm}
\begin{figure}[tbp]
$
\begin{array}{cc}
\epsfxsize=8cm \epsffile{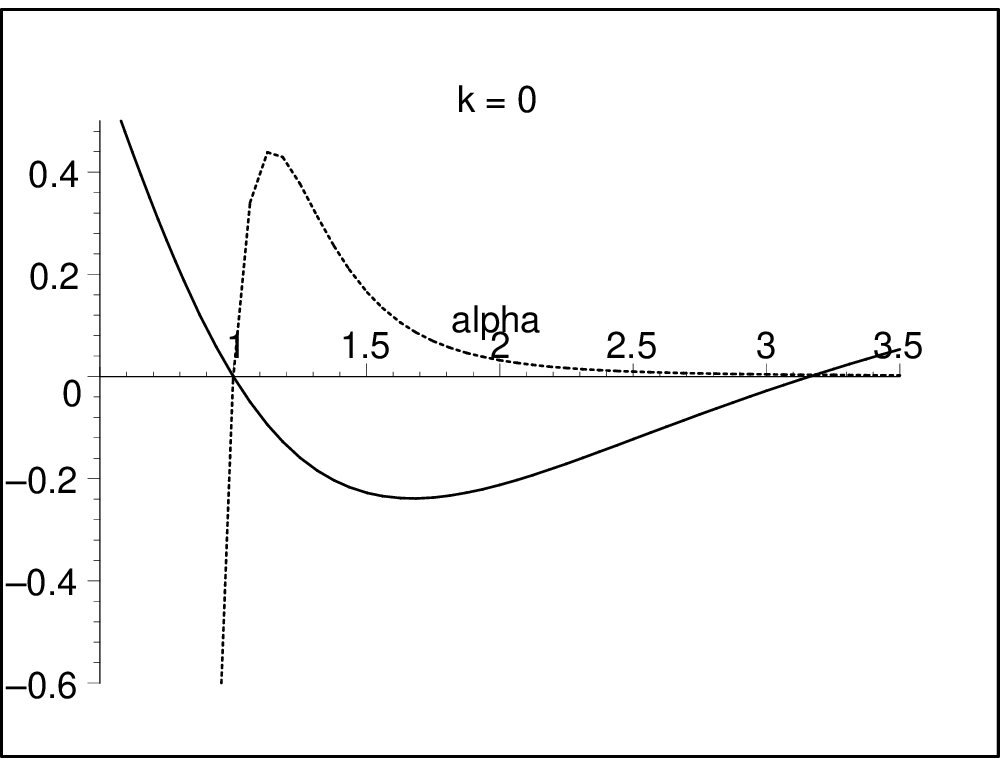} & \epsfxsize=8cm %
\epsffile{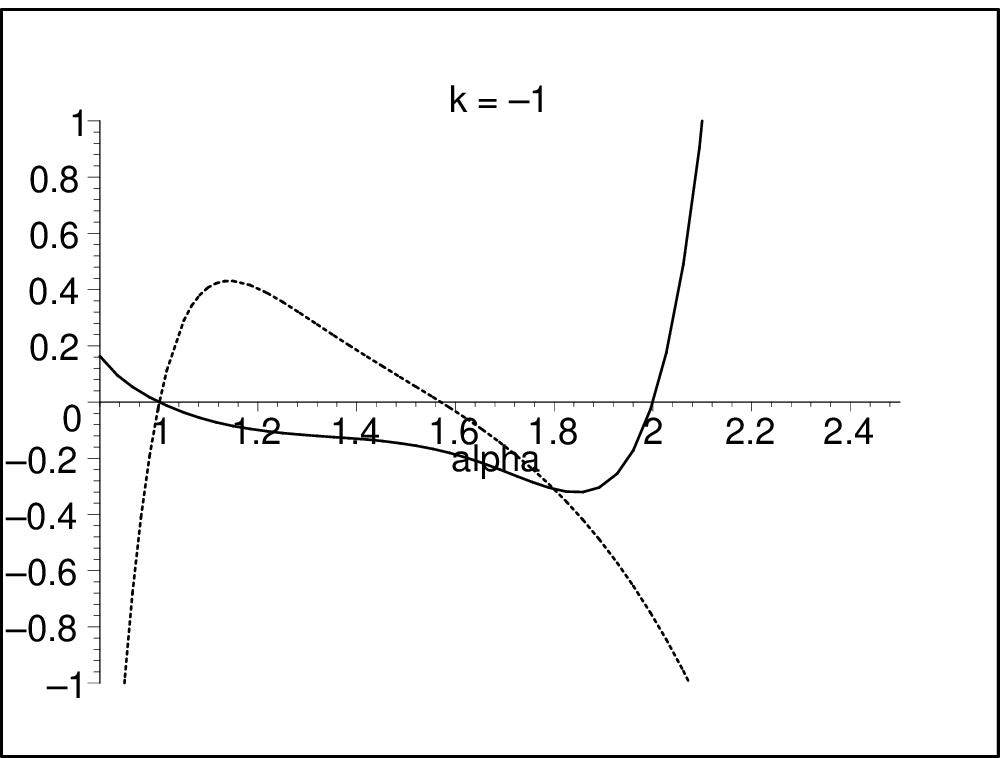}
\end{array}
$%
\caption{$10^{-6}(\partial ^{2}M/\partial S^{2})_{Q}$ (solid line) and $T$
(dashed line) versus $\protect\alpha $ for $b=0.2$, $n=4$, $l=1$ and $%
r_{+}=0.001$.}
\label{Fig3}
\end{figure}

\vspace{0.2cm}
\begin{figure}[tbp]
$
\begin{array}{cc}
\epsfxsize=8cm \epsffile{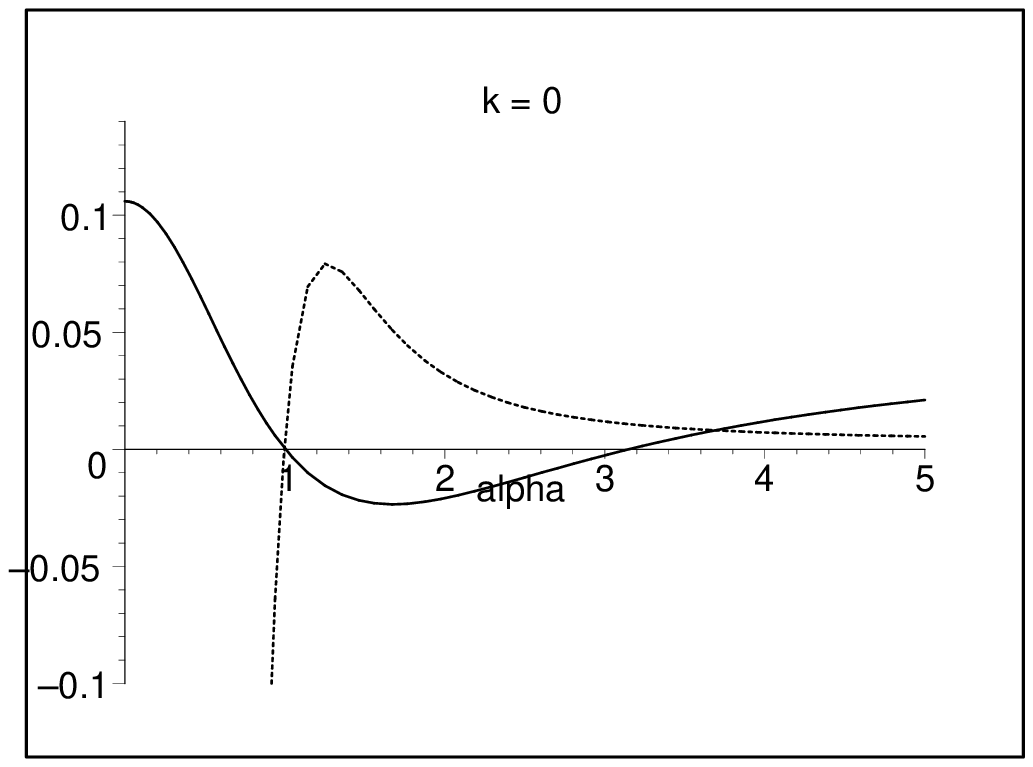} & \epsfxsize=8cm %
\epsffile{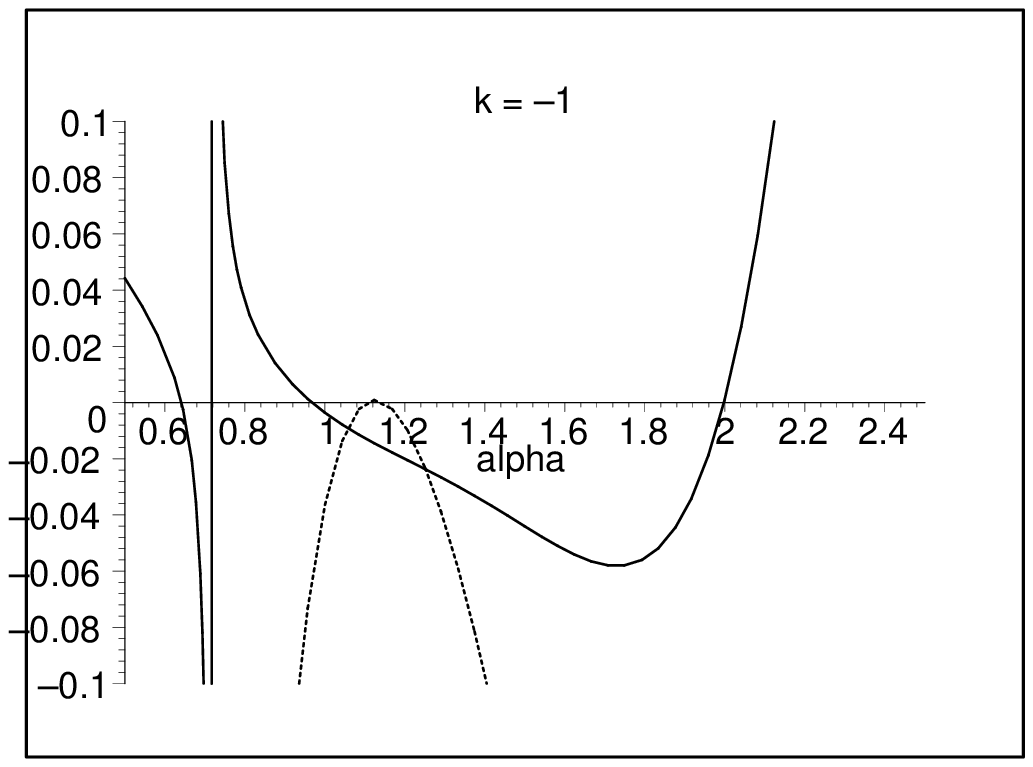}
\end{array}
$%
\caption{$10^{-5}(\partial ^{2}M/\partial S^{2})_{Q}$ (solid line) and $T$
(dashed line) versus $\protect\alpha $ for $b=0.2$, $n=4$, $l=1$ and $%
r_{+}=0.01$.}
\label{Fig4}
\end{figure}

\vspace{0.2cm}
\begin{figure}[tbp]
$
\begin{array}{cc}
\epsfxsize=8cm \epsffile{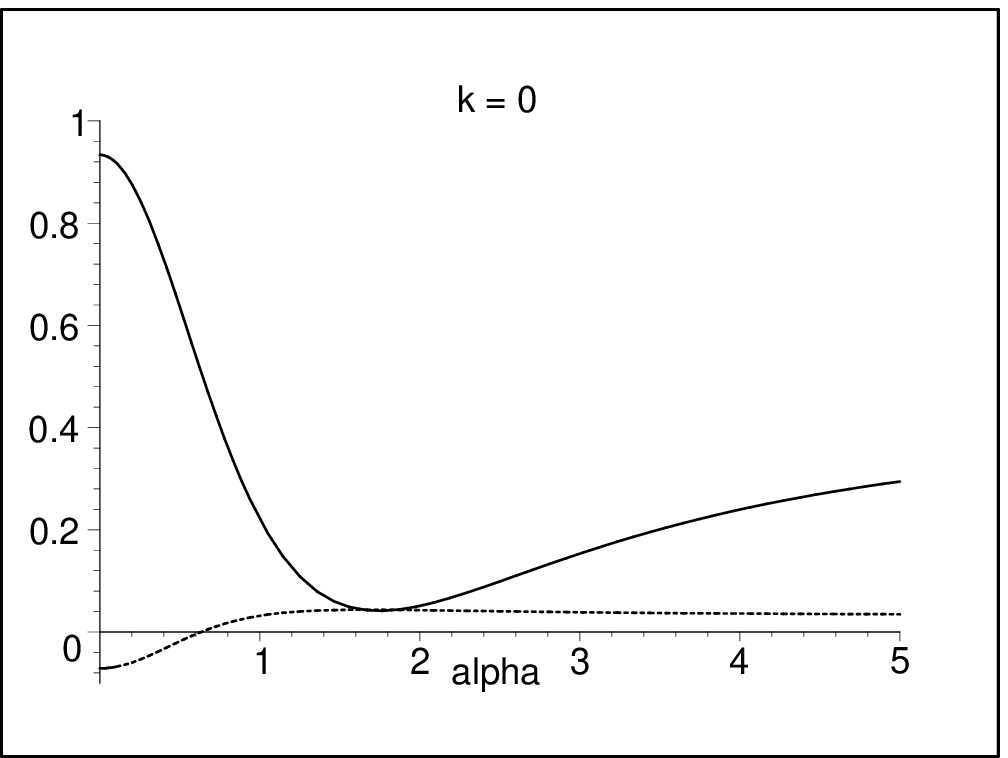} & \epsfxsize=8cm %
\epsffile{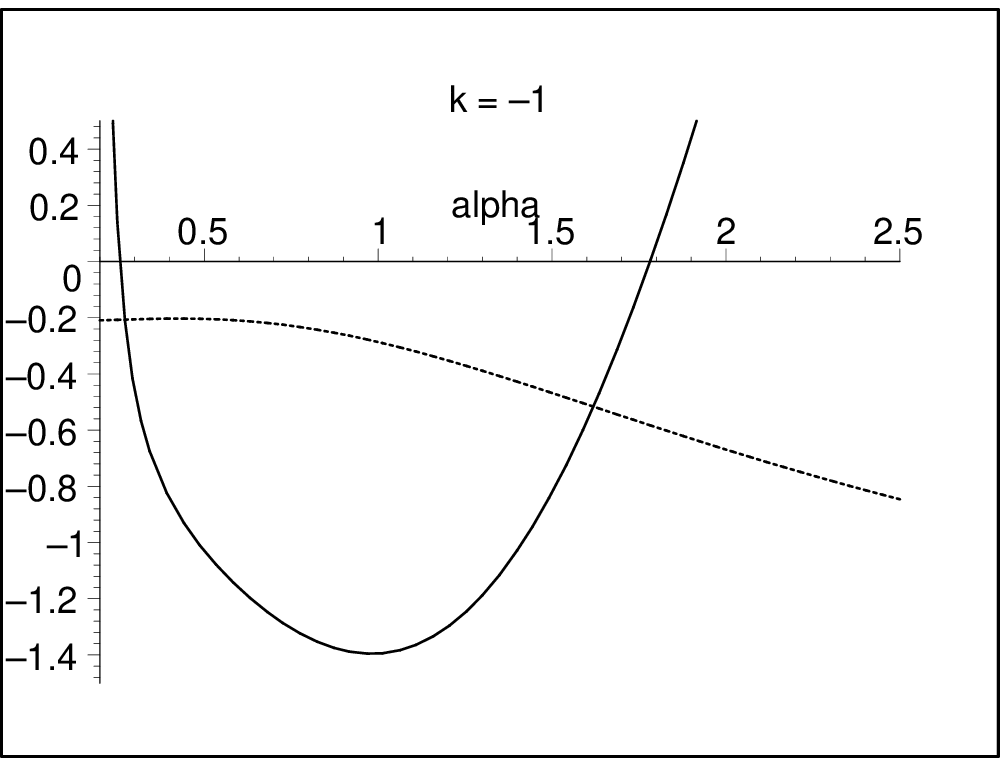}
\end{array}
$%
\caption{$10^{-2}(\partial ^{2}M/\partial S^{2})_{Q}$ (solid line) and $T$
(dashed line) versus $\protect\alpha $ for $b=0.2$, $n=4$, $l=1$ and $%
r_{+}=0.1$.}
\label{Fig5}
\end{figure}

 \vspace{0.2cm}
\begin{figure}[tbp]
$
\begin{array}{cc}
\epsfxsize=8cm \epsffile{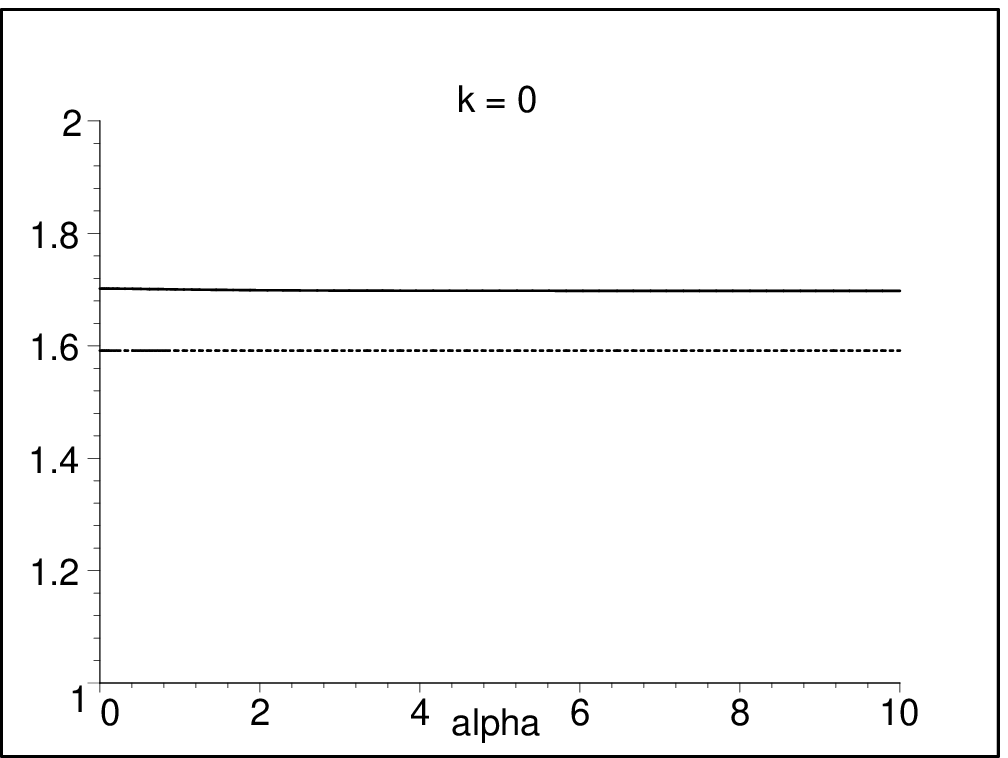} & \epsfxsize=8cm %
\epsffile{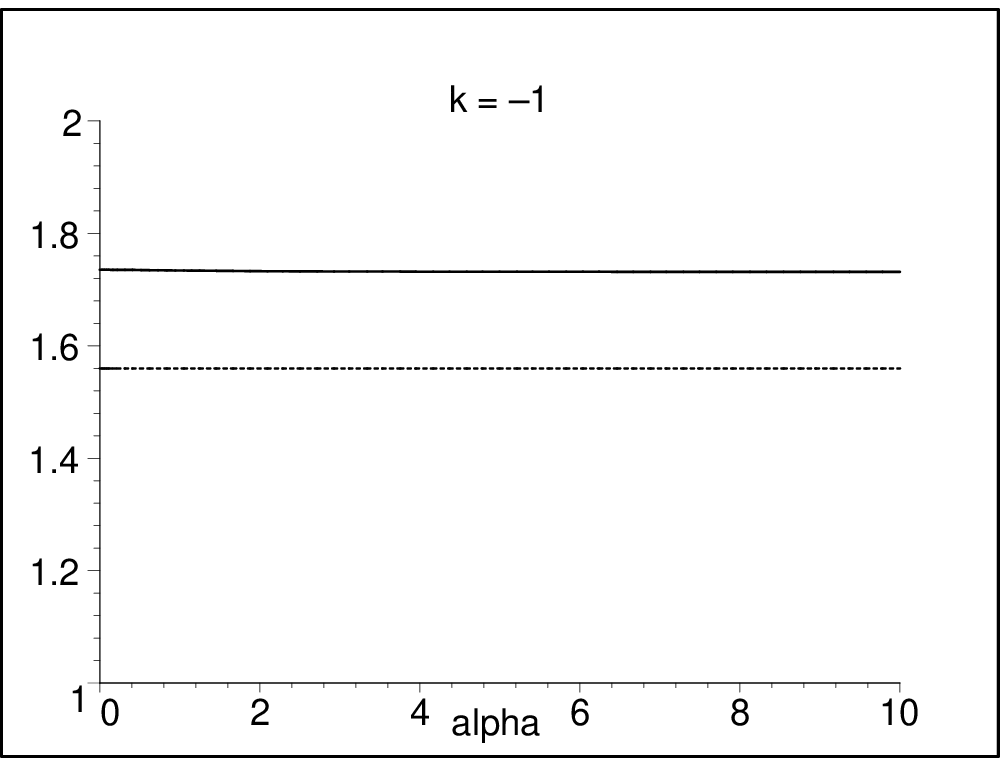}
\end{array}
$%
\caption{$10^{2}(\partial ^{2}M/\partial S^{2})_{Q}$ (solid line) and $T$
(dashed line) versus $\protect\alpha $ for $b=0.2$, $n=4$, $l=1$ and $%
r_{+}=5 $.}
\label{Fig6}
\end{figure}

\section{Stability in the canonical ensemble\label{Stab}}

Finally, we study the thermal stability of the solutions in the canonical
ensemble. In particular, we will see that the scalar dilaton field makes the
solution unstable. The stability of a thermodynamic system with respect to
small variations of the thermodynamic coordinates is usually performed by
analyzing the behavior of the entropy $S(M,Q)$ around the equilibrium. The
local stability in any ensemble requires that $S(M,Q)$ be a convex function
of the extensive variables or its Legendre transformation must be a concave
function of the intensive variables. The stability can also be studied by
the behavior of the energy $M(S,Q)$ which should be a convex function of its
extensive variable. Thus, the local stability can in principle be carried
out by finding the determinant of the Hessian matrix of $M(S,Q)$ with
respect to its extensive variables $X_{i}$, $\mathbf{H}_{X_{i}X_{j}}^{M}=[%
\partial ^{2}M/\partial X_{i}\partial X_{j}]$ \cite{Cal,Gub}. In our case
the mass $M$ is a function of entropy and charge. The number of
thermodynamic variables depends on the ensemble that is used. In the
canonical ensemble, the charge is a fixed parameter and therefore the
positivity of the $(\partial ^{2}M/\partial S^{2})_{Q}$ is sufficient to
ensure local stability. In Figs. \ref{Fig3} - \ref{Fig6} we show the
behavior of the $(\partial ^{2}M/\partial S^{2})_{Q}$ as a function of the
coupling constant parameter $\alpha $ for different values of the size of
black brane $r_{+}$. In order to investigate the stability of black branes,
we plot both $(\partial ^{2}M/\partial S^{2})_{Q}$\ and $T$ \ in one single
figure for various values of $r_{+}$\ or $\alpha $. Of course, one should
note that in these figures, only the positive values of temperature
associated to the event horizon of the black branes, and negative values of
temperature belong to inner horizon which we are not interested in. We
discuss these figures for $k=0$ and $k=-1$, separately.

\textbf{$k=0$}:

As we discussed in \ the last section, small black brane with a radius $%
r_{+} $\ exist when $\alpha >\alpha _{\mathrm{ext}}$. Figures \ref{Fig3} and
\ref{Fig4} show that these small black branes are stable provided $\alpha
>\alpha _{\mathrm{crit}}$. On the other side, large black branes are stable
as one may see in Figs. \ref{Fig5} and \ref{Fig6}.

\textbf{$k=-1$}:

Figures \ref{Fig3} and \ref{Fig4} show that small black branes
exist only for medium values of $\alpha $\ ($\alpha
_{1\mathrm{ext}}<\alpha <\alpha _{2\mathrm{ext}}$), but they are
unstable. On the other side, large black branes are stable as one
may see in Figs. \ref{Fig5} and \ref{Fig6}.
\begin{figure}[tbp]
\epsfxsize=8cm \centerline{\epsffile{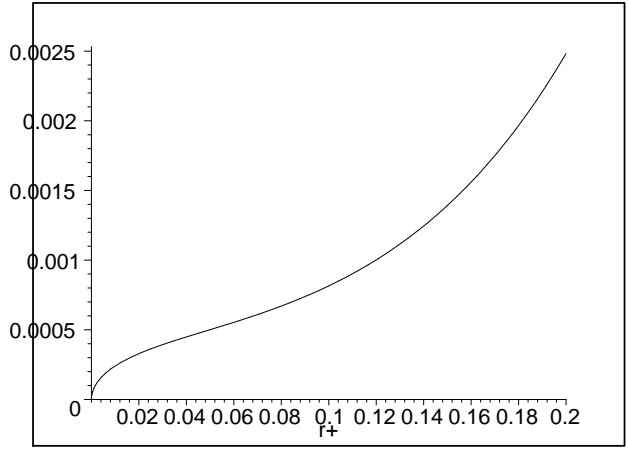}}
\caption{$F^{\mathrm{off}}$ versus $r_{+}$ for $b=0.2$, $n=4$,
$l=1$, $k=0$, $\protect\alpha =1.2$ and $T=0.42$.} \label{Fig1F}
\end{figure}

\begin{figure}[tbp]
\epsfxsize=8cm \centerline{\epsffile{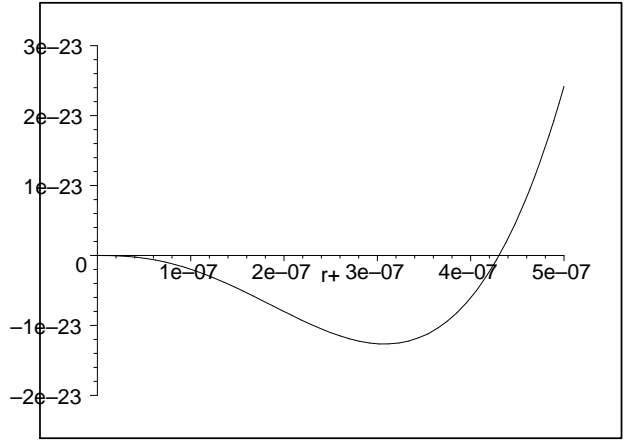}}
\caption{$F^{\mathrm{off}}$ versus $r_{+}$ for $b=0.2$, $n=4$,
$l=1$, $k=0$, $\protect\alpha =4$ and $T=0.0001$.} \label{Fig2F}
\end{figure}

In order to confirm the stability analysis of the black branes,
one can also use the generalized free energy \cite{Myung}
\[
F^{\mathrm{off}}(b,\alpha ,r_{+},T)=M(b,\alpha ,r_{+})-TS_{+}(b,\alpha
,r_{+}),
\]
which applies to any value of $r_{+}$ with a fixed
temperature $T$. This off-shell free energy reduces to the on-shell free
energy at $T=T_+$:
\[
F=M-T_{+}S_{+},
\]
which is the Legendre transform of $M$ with respect to $S$. As an
example of using the stability analysis by off-shell free energy,
we plot $F^{\mathrm{off}}$ versus $r_+$ for $k=0$, $\alpha=1.2$
and $\alpha=4$. These are plotted in Figs. \ref{Fig1F} and \ref{Fig2F},
which show that the black brane is unstable for $\alpha=1.2$,
while it is stable for $\alpha=4$. These figures confirm the
stability analysis of Fig. \ref{Fig4}. One may also confirm other
stability analysis given in this section.

\section{Conclusions\label{Sum}}
In $(n+1)$-dimensions, when the $(n-1)$-sphere of black hole event
horizons is replaced by an $(n-1)$-dimensional hypersurface with
zero or negative constant curvature, the black hole is referred to
as a topological black brane. The construction and analysis of
these exotic black branes in AdS space is a subject of much recent
interest. This is primarily due to their relevance for the AdS/CFT
correspondence. In this paper, we further generalized these exotic
black brane solutions by including a dilaton and the
electrodynamic fields in the action. We obtained a new class of $(n+1)$%
-dimensional $(n\geq 3)$ topological black brane solutions in
Einstein-Maxwell-dilaton gravity in the background of AdS spaces. Indeed,
the dilaton potential plays a crucial role in the existence of these black
brane solutions, as the negative cosmological constant does in the
Einstein-Maxwell theory. In the absence of a dilaton field ($\alpha =\gamma
=0$), our solutions reduce to the $(n+1)$-dimensional topological black brane
solutions presented in \cite{Cai33}. We computed the entropy, temperature,
charge, mass, and electric potential of the topological dilaton black branes
and found that these quantities satisfy the first law of thermodynamics.

For the thermodynamical analysis of the black branes, we divide
the black branes into three classes. The black brane is said to be
\textquotedblright small\textquotedblright\ when the radius of
event horizon is much smaller than the parameter $b$ , it is
called \textquotedblright medium\textquotedblright\ if $r_{+}$\
and $b$\ have the same order and it is
called \textquotedblright large\textquotedblright\ for large values of $%
r_{+} $\ with respect to $b$. The radius of event horizon is always larger
or equal to $r_{\mathrm{ext}}$\ for the black branes with flat horizon,
while it is in the range $r_{_{+}}<r_{1\mathrm{ext}}$\ or $r_{+}>r_{2\mathrm{%
ext}} $\ for hyperbolic black branes. We found that the existence of $r_{%
\mathrm{ext}}$\ for $k=0$, and $r_{1\mathrm{ext}}$\ for $k=-1$\ depends on
the value of $\alpha $. For the case of $k=0$\ with large values of $\alpha $%
\ ($\alpha >\alpha _{\mathrm{ext}}$), one can have black brane with any
size. This feature shows that dilaton change the thermodynamics of black
branes drastically. In this case, the temperature goes to infinity as $r_{+}$%
\ goes to zero for $n>4$, while it approaches to a constant for $n=4$. For
the case of $k=-1$, one may have small black branes provided $\alpha _{1%
\mathrm{ext}}<\alpha <\alpha _{2\mathrm{ext}}$, while large black branes
exist for any value of $\alpha$. Again, this is a drastic change in the
properties of the solutions because of the dilaton field. We analyzed the
thermal stability of the solutions in the canonical ensemble by finding a
Smarr-type formula and considering $(\partial ^{2}M/\partial S^{2})_{Q}$\
for the charged topological dilaton black brane solutions in $(n+1)$\
dimensions. We showed that for the case of $k=0$, small black branes are
unstable for $\alpha _{\mathrm{ext}}<\alpha <\alpha _{\mathrm{crit}}$, while
they are stable for $\alpha >\alpha _{\mathrm{crit}}$. Also, in this case
large black branes are stable for arbitrary $\alpha$. For the case of $k=-1$%
, small black branes are unstable, while large ones are stable. That is
dilaton with the potential given in Eq. (\ref{V(Phi)}), changes the
stability of the black branes drastically.

%%%%%%%%%%%%%%%%%%%%%%%%%%%%%%%%%%%%%%%%%%%%%%%%%%%%%%%%%%%%%%%%%%%%%%%
\acknowledgments{We thank the referees for constructive comments.
This work has been supported by Research Institute for Astronomy
and Astrophysics of Maragha, Iran.}

\end{document}